# On the derivation of wave function reduction from Schrödinger's equation: A model

Roland Omnès*
Laboratoire de Physique Théorique[#]
Université de Paris XI, Bâtiment 210, F-91405 Orsay Cedex, France

**Abstract**

The possibility of consistency between the basic quantum principles and reduction (wave function reduction) is reexamined. The mathematical description of an organized macroscopic device is constructed explicitly as a convenient tool for this investigation. A derivation of reduction from quantum mechanics is proposed on a specific example, using standard methods of statistical physics. Although these methods are valid only "for all practical purposes", arguments are given to ascribe an emerging status to reduction (or the uniqueness of physical reality), similar to the status of classical physics. Examination of measurements of the particles in an EPR pair by two space-like separated apparatuses shows that the model is consistent with the non-separable character of quantum mechanics, although both measurements are local. Quantitative estimates are given.

### 1. Introduction and outlook

The uniqueness of physical reality, technically expressed by wave function reduction or shortly reduction, stands as a challenge for an understanding of quantum mechanics [1-3]. As well known, the problem was clearly stated by Schrödinger when a quantum observable with only two eigenvectors $|1\rangle$ and $|2\rangle$ is measured [4]: Everything is reduced to essentials and one considers some pointer (a cat in Schrödinger's example) as representing the measuring apparatus $A$. This pointer stands at a neutral position $a_0$ in a state $|a_0\rangle$ before measurement. The apparatus is reliable and, when the state of the measured system $m$ is a vector $|j\rangle$ ($j = 1$ or 2), the quantum evolution of the system $A + m$ under unitary dynamics yields an outcome:

$$|j\rangle|a_0\rangle \to |j\rangle|a_j\rangle, \qquad (1.1)$$

When the initial state of *m* is however a superposition

$$|\psi\rangle = c_1|1\rangle + c_2|2\rangle, \tag{1.2}$$

the superposition principle predicts a final outcome:

$$|\psi\rangle|a_0\rangle \to c_1|1\rangle|a_1\rangle + c_2|2\rangle|a_2\rangle. \tag{1.3}$$

Schrödinger stressed the wide generality of this transfer of superposition from a microscopic to a macroscopic level, and drew its troublesome consequences. The main one was a no-go assertion: quantum dynamics cannot be consistent with a unique datum at the end of a measurement. When this uniqueness is imposed for any cause whatever, the linearity of unitary dynamics is broken.

The literature on this problem is enormous (see however [1] for classic papers and [3] for a recent evaluation). Many clever ways out have been proposed, except a rather trivial one apparently: Could it be that Schrödinger's no-go assertion was not sufficiently substantiated and the uniqueness of data is finally another example of the versatility and power of the quantum laws? This possibility was not forgotten however and two significant works, by Wigner and by Bassi-Ghirardi, came back to it and enforced Schrödinger's conclusion by turning it into what is frequently described as no-go theorems [5, 6].

Many problems are associated with measurements and some progress has been made for several of them. Decoherence theory, for instance, shed some light on the absence of macroscopic interferences, which is one aspect of Schrödinger's problem [7-12]. Consistent histories removed many logical paradoxes [13-15]. Explicit derivations of classical dynamics clarified the status of classical determinism [16-17]. These results were not fully consistent however for two reasons: They could not avoid completely the reduction problem and they were only valid "for all practical purposes "(fapp), and not fundamentally according to Bell's distinction [2]. They shared at any rate a remarkable character, which was their derivation from the principle of quantum mechanics, even if only "fapp". They raised accordingly a wider question: Could it be that the quantum principles contain in germ every element of their own interpretation? This assumption, which will be called here the hypothesis of consistency, has been the guiding line of the present work.

The main objections against it are due to the no-go theorems and to the "fapp" character of previous results along the same line. But is it sure that the foundations under these theorems are unquestionable, and is it obvious that a fundamental approach is necessarily better than one pretending only at a validity "fapp"? One will look first at the second point: There are many remarkable experiments, particularly in quantum optics, which look very much like quantum measurements, but show no reduction [11, 18]. Real macroscopic devices on the other hand show reduction in the unique result of an individual measurement. There are other significant differences between these two cases: A real apparatus is not only macroscopic and behaves classically, but it involves always a detection mechanism in which a quantum event is amplified to a macroscopic level (bringing out for instance sparks in a wire chamber or bubbles in a bubble chamber). This amplification process is irreversible and therefore exhibits the direction of time. But how could one include that in a fundamental approach, since the only indications about the direction of time are at best valid for practical purposes [12, 17, 19]?

Another difference is the existence of organization in a real measuring device and Laughlin stressed this character as incompatible with the fundamental quantum approach [20]. He asserted that the "emergence of objects" from quantum grounds through self-organization should be taken for granted, rather than using formally basic elementary quantum mechanic for highly complex systems in a rather vague way.

These questions are also significant for the assumptions of no-go theorem. They express a fundamental standpoint, but a satisfactory definition of such a standpoint is difficult. It can bring one into elaborate philosophical distinctions with different meanings for what is supposed to be "fundamental", "universal, or "real". Nevertheless, when dealing with the measurement problem, it seems that most of these approaches agree on representing a measuring apparatus by a wave function $\psi$ (or a state vector $|a\rangle$), evolving under the action of a unitary operator $U(t)$ [3]. Sometimes, $\psi$ is even supposed to represent a large part of the universe [6]. The meaning of such an absolute wave function (or an absolute density matrix) is not clear however and its existence is questionable, as discussed in Appendix A. One might notice on the other hand that this concept has not found other uses in physics than getting no-go assertions or somewhat metaphysical constructions [21], so that the main consequence of its reconsideration would only be to raise again the question of consistency.

When considering then the objection of the "fundamental" attitude against using a "fapp" approach for understanding the uniqueness of reality, the argument can be easily turned the other way round: If the hypothesis of consistency is valid, it implies that reduction is more or less an ordinary physical effect, leading to uniqueness under specific conditions. There is nothing implausible in such a possibility since the uniqueness of reality, as one can test it, holds mainly at macroscopic scales or in some properties of macroscopic objects. There is no sign of its validity at the atomic scale and rather contrary evidence from quantum theory. One may therefore expect that reduction has a restricted domain of action and uniqueness holds at a restricted scale. If so, reduction and uniqueness would not belong to the axiomatic center of quantum mechanics and would rather be an emerging property, like classical physics, irreversibility and organization. Accordingly, one should, and not only one might, consider the question of reduction within the scope of practical physics, with its successful though strictly practical methods

The present work will therefore assume an essential role of organization, as put forward by Laughlin. One cannot however be satisfied by Laughlin's standpoint as it was expressed, because of its incompleteness [20]: It does not explain for instance the mechanism through which the uniqueness of an object emerges and how a self-organized structure reacts when an outside influence would tend to bring it into a state of superposition. More significantly, it does not explain why a final datum is random and why the results of a series of identical measurements satisfy Born's probability rule. When one thinks moreover of two universal properties of a real measuring device, which are its capacity of amplifying a microscopic event and of registering it (sometimes, these two properties coincide), one is led to consider that an overall organization of this device is necessary. But whereas Laughlin stressed the role of self-organization, this is different: Self-organization is a powerful but somewhat fuzzy concept, and it is still directly associated with the quantum principles in the unique case of solid-state physics [22], which is why the present work will deal only with a pointer consisting of a moving solid.

Organization, in the usual sense of this word, is different from self-organization and rather obvious when used with the usual meaning of this word: One sees its significance for instance in a clock, which is an organized assembly of solid elements. It will be shown in Section 2 that it can be described by a mathematical formalism in which organization appears as a collection of conserved quantities. This is used in Section 3 to stress the significance of a change in organization when a quantum measurement is performed (because of irreversible events), and the fact that these changes are drastically different when the measured state is either $|1\rangle$ or $|2\rangle$ in Eq.(1.1), and again different in the case abstractly described by Eq.(1.3). These differences are so significant that they can imply fluctuations in the probabilities of the two channels, which become time-dependent and are no more exactly given by $p_1 = |c_1|^2$ and $p_2 = |c_2|^2$.

In order to appreciate the existence and meaning of these fluctuations, one selects in Section 4 a simple model for which the quantum laws can be used directly according to standard theoretical methods. Two parts of a measuring apparatus are privileged in this model: a reactive region *R* and a pointer *P*. In *R*, the measured system *m* creates some microscopic events, which are amplified through irreversible processes producing macroscopic signals. This process takes enough time to let decoherence act in *R* and separate the two signals, before they can act classically on the pointer and move it. When beginning its motion, the pointer is therefore already insensitive to the non-diagonal terms $|1\rangle\langle 2|$ and $|2\rangle\langle 1|$ in the state of *m*.

The pointer *P* is supposed solid and moving, so that its final position in space will indicate the result of the measurement. This example of a mechanically moving pointer is rather special and it was chosen only for its tractability. The pointer is supposed near thermal equilibrium at a temperature around room temperature and its state is described according to the methods statistical physics and particularly the use of Gibbs subsystems. This approach assumes necessarily that different subsystems, *i.e.*, different small regions *β* and *β'* inside *P*, have incoherent quantum states. This is certainly the main assumption of the present approach, which is also intrinsic by the way to the existence of entropy [23], and its relation with information [24], but can be only valid "fapp". It is exactly opposite to the usual fundamental approach [3-6], in which the wave function of the whole apparatus, and perhaps of a large part of its environment, is supposed rigidly coherent. Such an opposition in approaches raises obviously difficult questions regarding the interpretation of quantum states, briefly discussed in Appendix A.

When the signals associated with the *m*-states $|1\rangle\langle 1|$ and $|2\rangle\langle 2|$ act on the pointer, they move it differently and the quantum state of a Gibbs subsystem *β* in the pointer splits into two different subsystems *β*1 and *β*2. The wave functions in *β*1 and *β*2 do not become immediately orthogonal when the signals begin to act on the pointer, and one can show that full orthogonality between all the wave functions belonging to *β*1 or to *β*2 is not reached until the positions $x_1$ and $x_2$ of the pointer differ by more than about $10^{-11}, 10^{-10}$ cm. This period will be called the proximity stage and occurs just at the beginning of registration by the pointer. During this period, an atom in a wave function entangled with $|1\rangle$ in *β*1 can interact with other atoms in a state entangled with $|2\rangle$ in *β*2 and this interaction yields fluctuations in the channel probabilities $p_1$ and $p_2$. One can compute explicitly these fluctuations after a canonical transformation from atomic observables to phonon observables, and they are found strong.

The transition from fluctuations to reduction can be finally derived, using previous results by Pearle [25], whose basic assumptions are found valid in the present case (without appealing to nonlinear Schrödinger equations, as Pearle supposed necessary). One finds quantitatively that reduction is a very rapid process occurring during the proximity stage in the present model and satisfying the usual assumptions for the reduction of a wave function.

Section 5 is devoted to an application of the theory to the measurement of an EPR pair by two apparatuses with a spacelike separation: the results agree with experiment [26-28].

## 2. Mathematical description of organized systems

*Organization and its mathematical framework*

The notion of organization is wide and well known, but it will be convenient to recall its meaning on a simple example, such as an ordinary mechanical clock. It is made of various pieces, wheels, springs, hands and so on, which can be distinguished by a label $\beta$. (not to be confused with the label for a Gibbs subsystem in the introduction). These pieces have been manufactured and assembled in a careful way and the constraints resulting from their links and contacts are so compelling that they leave only a finite number of degrees of freedom, mathematically described by Lagrange-Hamilton coordinates $q_r$. In the present work, organization will be used to denote these characters and properties in various pieces of a physical system, together with a permanence of their assembly. Organization is essential in both the natural world and a laboratory, and it was formalized only in classical dynamics by Lagrange and Hamilton. Little use has been made of it however in quantum mechanics, except in terms of boundary conditions.

From the standpoint of quantum mechanics, every piece $\beta$ of the clock consists of atoms in a solid state. Its composition, crystal phase, lattice, and mechanical properties are governed in principle by quantum physics and, more specifically, by solid-state physics. Its shape requires however special attention: It originated in the past when $\beta$ was manufactured and was conserved later on. This permanence provides therefore in some sense a conservation law. One will denote by $\sigma_\beta$ this boundary of $\beta$. It exists for all practical purposes, but it is not a sharp surface and, even when using a scanning tunneling microscope, one cannot define it below the Angstrom scale.

The relation of stable boundaries with conservation laws can be shown explicitly at the classical level: One considers a classical mark $M$ on $\sigma_\beta$ or inside the piece $\beta$ and one looks at its dynamical description: The configuration space $C$ of the clock with its coordinates $q_r$ defines completely the location and orientation of every piece $\beta$. A motion of the clock from a configuration $q_{r0}$ of the clock at time 0 to a configuration $q_r(t)$ acts on $\beta$ as a space displacement involving a space translation $a_\beta(q_r)$ and a rotation of axes $R_\beta(q_r)$. From the space position $x(t)$ of $M$ at time $t$, one can define another dynamical variable $x_0(t)$ through

$$x_0(t) \equiv R_\beta^{-1}\{q_r(t)\}[x(t) - a_\beta\{q_r(t)\}]. \qquad (1.1)$$

The permanence of shapes and organization means then that $x_0(t)$ is a constant of motion, which can be expressed by an equation $\{x_0, H\} = 0$ involving a Poisson bracket with the Hamilton function $H(q_r, p_r)$.

In that sense, the rigidity of every piece of the clock and the permanence of their organization would appear as a huge set of conserved quantities, as many as atoms in the clock if one used a classical description of these atoms. In quantum mechanics however, the boundaries are fuzzy and the atoms move around their lattice sites so that these constants of motion have a different expression. They can be replaced by a unique conservation law expressing altogether the self-organization of every piece $\beta$ and the overall organization, although this is only approximate (valid "fapp") To do so, one can still use the previous boundaries $\sigma_\beta$, but also take their fuzziness into account and turn it into a convenience: One considers a surface $\sigma'_\beta$ enclosing $\beta$ and remaining everywhere at a distance of the order of 1 Angstrom from its atoms, to make sure that these atoms stand well inside the domain $D_\beta$ with boundary $\sigma'_\beta$, up to small errors. An apparent inconvenience is that two such domains $D_\beta$ and $D_{\beta'}$ overlap when two pieces $\beta$ and $\beta'$ are in contact, but that will be reconsidered.

One can consider again the time evolution of the Lagrange observables as expressed with a good approximation by classical dynamics (using for instance microlocal analysis [17]). One can then use the previous geometrical quantities $a_\beta(q_r)$ and $R_\beta(q_r)$ and the analogue of the set of conserved quantities (1.1) for all the atoms in $\beta$ with position coordinates $x_k$ by means of a projection operator

$$F_\beta = \prod_{k \in \beta} \left[ \int_{\sigma'_\beta} U |x_k\rangle dx_k \langle x_k| U^{-1} \right] \tag{1.2}$$

where

$$U|x_k\rangle = \left| R_\beta^{-1}(q_r)\{x_k - a_\beta(q_r)\} \right\rangle. \tag{1.3}$$

Conservation of the shape of a piece $\beta$ amounts then to the set of approximate conservation laws

$$[H, F_\beta] = 0, \tag{1.4}$$

where $H$ is the basic Hamiltonian of the atoms in the clock. The permanence of organization amounts then to a global conservation law $[H, F] = 0$ for the projection operator

$$F = \prod_\beta F_\beta. \tag{1.5}$$

If one denotes by $E$ the basic Hilbert space representing the quantum states of atoms belonging to the clock, the set of these states showing organization is strongly constrained. This constraint restricts the states of the organized clock to the subspace $E' = F E$ of $E$. Similarly, if $H$ denotes the basic Hamiltonian of atomic particles, it reduces to the operator $H' = F H F$ when acting on the states of the clock. Since $H$ is local, the same is true of $H'$ as far as direct interactions between atoms are concerned. Because of this locality, $H'$ can take care of the solid-state structure of every component $\beta$, and of the microscopic details of a contact between two neighboring components $\beta$ and $\beta'$, including friction effects. The trick of using a loose boundary $\sigma_\beta$ is especially convenient in this regard, since it insures that the Hilbert space $E'$ is wide enough to represent the clock for all practical purposes.

*The state of an organized system*

When looking at a density matrix $\rho$ describing a state of the clock, one sees that the self-organization of every piece $\beta$ and the organization of the clock are respectively expressed by the relations

$$Tr(\rho F_\beta) \approx 1, \qquad Tr(\rho F) \approx 1. \qquad (1.6)$$

But a non-trivial question asks what is the exact meaning of this density matrix, or more generally the state of an organized system. Similar questions always occur in the foundations of quantum mechanics and are a prerequisite in some proofs of no-go theorems. The present approach presumed that a classical knowledge (or understanding) of the clock and its components would be sufficient as a starting point, but that does not imply an objective definition of $\rho$ and only restricts its mathematical framework. Conversely, one could say that in quantum mechanics, nothing but $\rho$ itself contains enough information for expressing the constraints of organization and constructing the mathematical framework ($E'$, $H'$). This significant point can be seen as follows:

The clock is made of atoms and one can use the set of position coordinates $\{x_k\}$ of these atoms as a complete set of commuting observables (leaving aside spin and other atomic quantum numbers). The matrix elements of $\rho$ are then functions $\rho(\{x_k\}.\{x'_j\})$ satisfying Bose-Einstein or Fermi-Dirac symmetries for identical atoms. If one selects one variable $x_1$, associated with the position of a definite atom (for instance a Cu atom), one can trace out all the other variables and obtain a one-atom density matrix $\rho_1(x_1, x'_1)$. It shows the distribution of all the Cu atoms everywhere in the system, because of the symmetries od identical particles. Its diagonal elements $\rho_1(x_1, x_1)$ provide a three-dimensional radiography of the Cu atoms in the clock, and similar mathematical operations on every kind of atom provide many views of the clock showing the shape of its components, their bounding surfaces, as well as the various crystal lattices with their geometric symmetries, cell size and orientation, together with dislocations and many other fine details such as chemical composition in various places. When this description is extended from the level of atoms to the deeper level of nuclei and electrons, one gets views of the covalent binding of atoms, the detailed structure of molecules, and so on. One can therefore consider organization as a set of mathematical structural properties *belonging to the density matrix*, which remain valid for all practical purposes during a long time. That does not say anything however about what could be *the* exact $\rho$, what it means and whether it exists, these questions being left for reconsideration in Appendix A.

## 3. Measurements as changes in organization

One considers now quantum measurements and defines a *real* measurement as involving an organized apparatus. This restriction does not mean of course that many remarkable experiments, particularly in quantum optics when the systems had no organization [18], dealt with unreal events. It means that real measurements are events where a unique reality comes out, for all practical purposes, whatever the underlying mechanism and perhaps because reality itself is essentially organized.

One will take an example where the measuring apparatus $A$ consists of three parts, denoted by $P$, $R$ and $S$. $P$ is a solid moving pointer with a position given by a unique Lagrange observable $X$. $R$ is a reactive region, which one can suppose unorganized (like a gaseous dielectric in a Geiger counter for instance). $S$ is the structure or support including for instance the walls of $R$, a ruler on which the pointer can slide or an axis around which it can rotate. $S$ can also insure an unstable sensitivity of $R$ to the action of the measured system (generating for instance an electric field in the dielectric of a counter). Generally, $S$ is not stationary and carries its own Lagrange observables. Before measurement, $A$ is an organized system, associated with a couple ($E'$, $H'$).

A real measurement amplifies a quantum event to a macroscopic scale, with a significant effect on the pointer position. When the measured system $m$ is for instance a charged particle entering into $R$, the amplification process goes through an ionization cascade where $m$ ionizes the medium, producing ions and primary electrons, these electrons are accelerated by the electric field and produce more ions together with secondary electrons, and so on. One can formalize this example in the following way: One assumes that, under the influence of either $|1\rangle$ or $|2\rangle$, two distinct tracks are produced in $R$, or in two distinct regions of $R$ are affected. One can then characterize them by the numbers of ions $Q_1$ and $Q_2$ in these regions and this example is especially convenient because the eigenvalues $q_1$ and $q_2$ of $Q_1$ and $Q_2$ are cleanly distinct integers.

To say in these conditions that a quantum measurement involves a change in organization does not suppose a meaning of organization that would be restricted to an assembly of self-organized components. In the present example, it means that new macroscopic observables, associated with irreversible signals, are generated and strongly affect the dynamics of an organized apparatus. In that sense, no self-organized component is structurally modified, no more than their geometric organization, but new forces appear and bring out new parameters affecting classical dynamics. The occurrence of these new forces would be enough, in classical physics, to define new formal dynamical systems. In the present case, it stands as the simplest expression of striking irreversible effects in the reactive region, which are enough to deserve a reconsideration of the mathematical framework of organization.

Although new self-organized elements can be produced in the reactive region $R$ (for instance droplets in a Wilson or bubbles in a bubble chamber), one will only assume in the present model that the atoms, ions and electrons in $R$ remain separated from $P$ and $S$. This condition can be conveniently expressed through the introduction of three orthogonal Hilbert spaces for the states of the reactive region, differing by the values of $q_1$ and $q_2$, namely a space $E_{R0}$ (corresponding to $q_1 = q_2 = 0$), $E_{R1}$ ($q_1 > 0$, $q_2 = 0$) and $E_{R2}$ ($q_1 = 0$, $q_2 > 0$). Four different Hilbert spaces are then necessary for a mathematical description of measurements, namely

$$E_0 \equiv E' = E_S \otimes E_P \otimes E_{R0}, \quad E_1 = E_S \otimes E_P \otimes E_{R1}$$

$$E_2 = E_S \otimes E_P \otimes E_{R2}, \tag{3.1}$$

$$E_{12} = E_1 \oplus E_2. \tag{3.2}$$

$E_0$ describes apparatus $A$ before measurement, $E_1$ or $E_2$ describe it when either state $|1\rangle\langle 1|$ or $|2\rangle\langle 2|$ is measured. These Hilbert spaces are orthogonal because of the orthogonality of $E_{R0}$, $E_{R1}$ and $E_{R2}$, but the tensor products in Eqs.(3.1) show explicitly the permanence of $S$ and $R$. The direct sum $E_{12}$ is relevant when a superposed state $|\psi\rangle\langle\psi|$ is measured. The Hilbert spaces $E_0$, $E_1$ and $E_2$ are related to the basic Hilbert space $E$ describing particles by different projection operators so that

$$E_0 = F'E, \quad E_1 = F_1 E, \quad E_2 = F_2 E. \tag{3.3}$$

It may be noticed that all the properties of orthogonality are carried by the states of the reactive region $R$, whereas the Hilbert spaces of $S$ and $R$ remain the same with the same organization. From this mathematical standpoint, a measurement appears as the occurrence of new effects expressing the production of macroscopic signals $q_1$ and $q_2$.

When acting on the Hilbert space $E_{12}$, the Hamiltonian is represented by a $2\times 2$ matrix with diagonal elements ($H_1, H_2$) and non-diagonal elements ($H_{12}, H_{21}$), where

$$H_1 = F_1 H F_1, \quad H_2 = F_2 H F_2, \quad H_{12} = F_1 H F_2, \quad H_{21} = F_2 H F_1. \tag{3.4}$$

One can then write down the density matrix of the system $A + m$ during the measurement process when $|1\rangle$ and $|2\rangle$ are taken as a basis for the Hilbert space $E_m$ of the measured system $m$. One thus has

$$\rho_{A+m} = \rho_1 |1\rangle\langle 1| + \rho_{12}|1\rangle\langle 2| + \rho_{21}|2\rangle\langle 1| + \rho_2 |2\rangle\langle 2|. \tag{3.5}$$

where $\rho_1$ and $\rho_2$ are self-adjoint, positive, with respective traces $p_1$ and $p_2$ such that $p_1 + p_2 = 1$, whereas $\rho_{12}$ and $\rho_{21}$ are adjoint of each other. At time zero, just before measurement, when the initial state of $m$ is $|\psi\rangle\langle\psi|$, with $|\psi\rangle = c_1|1\rangle + c_2|2\rangle$, one has

$$p_1(0) = |c_1|^2, \quad p_2(0) = |c_2|^2. \tag{3.6}$$

Because of entanglement of a state $|j\rangle$ of $m$ with states of $R$ belonging to $E_{Rj}$, the Schrödinger-Von Neumann equation becomes

$$id\rho_1/dt = [H_1, \rho_1] + H_{12}\rho_{21} - \rho_{12}H_{21}, \tag{3.7a}$$
$$id\rho_2/dt = [H_2, \rho_2] + H_{21}\rho_{12} - \rho_{21}H_{12}, \tag{3.7b}$$
$$id\rho_{12}/dt = H_1\rho_{12} - \rho_{12}H_2 + H_{12}\rho_2 - \rho_1 H_{12}, \tag{3.7c}$$

Taking traces, these equations yield

$$d\,p_1/dt = 2\,Im\,\{Tr(H_{12}\rho_{21})\} = -\,d\,p_2/dt \tag{3.8}.$$

These variations of quantum probabilities (or rather squared amplitudes) are not unfamiliar and well known in the case of chemical reactions or when two quantum states are linked through a non-diagonal coupling. Here however, the variations affect the probabilities of measurement channels, which were always found constant when organization was not taken into account . No such variations result moreover from decoherence (which is not linked with organization [11]) and this is why some authors incline to believe that decoherence could be consistent with Everett's many-worlds interpretation of quantum mechanics [12, 29]. It is especially remarkable that little attention was paid to the drastic difference between $H_1$ and $H_2$, which can be seen with a naked eye when a measuring device is working. $H_1$ and $H_2$ were always thought of as identical with the basic atomic Hamiltonian $H$, whereas the sharp difference between them is expressed as two physically different mathematical representations of $H$ in the present framework. Nevertheless, Eq.(3.8) is not obvious. The existence of $H_{12}$, its meaning and its action remain to be understood, and that will require a closer analysis.

## 4. A model of reduction

*Background of a model*

This analysis will go through several steps, some of which delicate or in plain opposition to a fundamental approach excluding in principle any "fapp" consideration. This means that one takes definitely the standpoint of a reduction mechanism, or more generally of the uniqueness of reality, as a macroscopic property conditioning every practical aspect of physics, like the classical representation of macroscopic properties of nature is a condition for giving a meaning to phenomena [3]. It may be that this is closely linked with the wide and somewhat fuzzy notion of self-organization, but rather than envisioning a "different universe" [20], one must remain within the framework of standard quantum mechanics to justify the hypothesis of consistency. This is why one will not try to look for reduction in a reactive region where irreversible processes occur, since they are too complex, but one will look at what happens in a solid pointer where solid-state physics can be used. Moreover, one will not consider the electric phenomena occurring almost always in an apparatus, but one will deal with a much less general case where the pointer moves mechanically, either sliding along a support or turning around an axis. The reason for this choice is entirely practical, namely because it is easier.

Let one begin with one of the questions at the end of the previous section: the meaning of $H_{12}$ As stressed by Schrödinger [4], the essence of the measurement problem is entanglement between the states of the measuring apparatus $A$ with the states of the measured system $m$. The meaning of $H_{12}$ is then rather obvious in the present model: It arises from the potential interactions between atoms where an atom, in a state of $A$ entangled with $|1\rangle$, interacts with another atom in the state entangled with $|2\rangle$.

Let one then consider decoherence in $R$. A pointer initially at rest is not free in a real device,. External forces from the signals in $R$ must be large enough to put it into motion and break its resistance to motion, due for instance to friction. This means that the signals $Q_1$ and $Q_2$ must be large enough and in practice already classical before they can exert an action on the pointer. Their growth to that level requires a finite time and, since they are macroscopic and sufficiently distinct to bring out different registrations, the standard conditions for decoherence in *R* are satisfied. [12].

Ignoring for a moment the support and the pointer and concentrating on the density matrix for the reactive region *R* and the measured system *m*. Writing it in the form (3.5), one has

$$\rho_{R+m} = \rho_{R1}|1\rangle\langle 1| + \rho_{R12}|1\rangle\langle 2| + \rho_{21R}|2\rangle\langle 1| + \rho_{R2}|2\rangle\langle 2|. \quad (4.1)$$

Decoherence can take many aspects, but it amounts generally to an interaction involving the macroscopic quantities $Q_1$ and $Q_2$ and a large number of random observables defining an environment *e*. An external system such as the pointer, which interacts with $Q_1$ and $Q_2$ and not directly with *m*, which is moreover practically insensitive to the subsystem *e* of *R*, can experience only an action of the reduced density matrix

$$\rho_{R+m}^{(red)} = Tr_e(\rho_{R+m}). \quad (4.2)$$

According to decoherence theory [12], the non-diagonal elements of this matrix vanish very rapifly and one has for instance:

$$\rho_{R12}^{(red)}(t) \approx \exp(t/t_{deco})\rho_{R12}^{(red)}(0). \quad (4.3)$$

where $t_{deco}$ is a very short decoherence time. The second and third terms in the right-hand side of Eq.(4.1) become therefore rapidly negligible.

Before looking at the consequences of the resulting diagonal form, one may remember that there exist objections against the use of decoherence in a theory of measurements [30, 3]. They rely essentially on the idea that, if the whole system is described by a wave function, partial traces cannot be invoked to circumvent that property. As a matter of fact, this is the root of all the no-go assertions and theorems since Schrödinger [1-6], and it is examined in Appendix A. Some previously neglected aspects of this problem are noticed there and they will be considered sufficient in the present section for pursuing investigations.

*Some consequences of statistical physics*

The significance of decoherence appears more clearly when one applies statistical physics to the description of a measurement. Let one use for that purpose the notion of Gibbs ensembles and of their standard quantum expression to the system *A + m*. Considering *R* and *P* as two subsystems and disregarding *S* for convenience, one can write the density matrix of *A + m* at time *t* as

$$\rho = p_1\rho(R_1) \otimes \rho(P_1) \otimes |1\rangle\langle 1| + p_2\rho(R_2) \otimes \rho(P_2) \otimes |2\rangle\langle 2|. \quad (4.4)$$

The various matrices denoted by $\rho$ in the right-hand side are positive with trace 1, $p_1$ and $p_2$ being the probabilities of the two channels at time *t*. There are no terms $|1\rangle\langle 2|$ or $|2\rangle\langle 1|$ in this equation after decoherence, because of the previous argument: If such non-diagonal terms were written, one would have to take their partial trace over the internal environment *e* of *R* in order to exhibit the classical action of the signals on the pointer, and that would suppress non-diagonal terms.

The next step consists in applying statistical physics to the pointer, which is supposed not far from thermal equilibrium. It is then decomposed as usual into subsystems, which will be indexed by a label $\beta$ (not to be confused with the notation in Section 2). Eq.(4.4) becomes then

$$\rho = p_1 \rho(R_1) \otimes \bigotimes_{\beta} \rho_{\beta 1} \otimes |1\rangle\langle 1| + p_2 \rho(R_2) \otimes \bigotimes_{\beta} \rho_{\beta 2} \otimes |2\rangle\langle 2|. \qquad (4.5)$$

This expression, like Eq.(4.4), is of course approximate and valid only "fapp". It can be justified in various ways, one of them being to perform an averaging of the density matrix over a short time when introducing for instance entropy [23] The Hamiltonian of the pointer is then written as a sum

$$H = \sum_{\beta} H_{\beta} + \sum_{\beta\beta'} H_{\beta\beta'}, \qquad (4.6)$$

including the energy of subsystems and their interactions. Eq.(4.5) is then valid over a time interval during which energy exchanges between different subsystems are negligible, "fapp". The strong metastability of $R$ with its exponentially growing signals cannot obviously be treated in the same way, which is why this region is considered here as a separate subsystem with no internal decomposition.

Eq.(4.5) has a remarkable consequence however. When looking at the eigenvectors of $\rho$, one sees that they involve a sum of terms consisting of the product of either $|1\rangle$ or $|2\rangle$ by a wave function of $R$, and products of independent wave functions $\psi_{\beta}$ for the subsystems $\beta$. This product of functions $\psi_{\beta}$ means that the approximation treats them as independent and incoherent (with no phase relation). In that sense, Eq.(4.5) is in opposition with the fundamental approach where the whole system is supposed described by a unique wave function $\psi$ with inseparable phase correlations [3, 6]. On the other hand, such a rigid conception of $\psi$, which is at the basis of no-go theorems, is questionable as discussed in Appendix A. In any case, this point belongs to the delicate ones that were mentioned and it will be left aside presently.

Another aspect of the same question about incoherence is that one may wonder about the lower limit in the size of the subsystems for which Eq.(4.5) is valid. This equation is usually applied to irreversible processes not far from thermal equilibrium, for which the exact size of $\beta$ is irrelevant, but in the present case, one would need to know how far one can go in reducing this size and using incoherent $\psi_{\beta}$'s together with an estimate of errors. These questions seem absent from the literature, perhaps because there were no occasions to use Gibbs ensembles and superposition together, but it is fair to say that the tentative answer that will be given later to this question will be mainly a guess.

*The notion of proximity*

Another notion entering in the present calculations can be introduced by looking at Wigner's fundamental paper, in which he discarded consistency between reduction and Schrödinger's equation [5]. His argument relied on the orthogonality of two possible final states of the pointer, but one may wonder what happens when the pointer is just beginning to move. Are its two states orthogonal? For the sake of orientation, let one suppose that the pointer slides along the *x*-direction. The matrices $\rho(P1)$ and $\rho(P2)$ in Eq.(4.4) are very similar, and one may consider for illustration two eigenvectors $|x_1\rangle$ and $|x_2\rangle$, of these matrices, where $x_1$ and $x_2$ denote the positions of the pointer. Introducing $\xi = x_1 - x_2$, one shows in Appendix B that these two states are not orthogonal for small values of $\xi$ and one has only the inequality:

$$|\langle x_1|x_2\rangle|^2 \leq \exp(-N\xi^2/4\alpha\Delta^2). \qquad , \qquad (4.7)$$

where $\Delta$ denotes the quantum uncertainty for the displacement of an atom in its ground state from its lattice site, $N$ is the number of atoms in the pointer and the coefficient $\alpha$ representing the effect of thermal motion on the displacements of atoms depends on the temperature and is always larger than 1 though of order O(1). Equality holds only in this relation when the two states $|x_1\rangle$ and $|x_2\rangle$ are identical, except for their relative displacement $\xi$. Remembering however that typical values for $\Delta$ are $10^{-9}$ cm and that $N$ can be as large as the Avogadro number, the displacement $\xi$ would have to be less than $10^{-20}$ cm to yield non-orthogonal states. It would seem therefore that the situation of proximity is so extreme that it would have no place in physics.

This is where incoherence in the wave functions enters when one considers a subsystem β, because one must then replace $N$ by $N_\beta$. One will encounter later on relevant values of $N_\beta$ of the order of $10^5$ and $\xi$-values of the order of $10^{-11}$ cm. This is still quite small, but the relevant parameter is not the distance. It is the time necessary for the pointer to move on that distance when starting from position 0, when compared on one hand to the rate of fluctuations in the channel probabilities and, on the other hand, with the time of growth of the signals. It will be found that the rate of fluctuations is so high and the time of growth of signals long enough for yielding strong effects implying reduction during the proximity stage, at least in the present model.

*Fluctuations in channel probabilities*

One may come at last to the question of existence of fluctuations in the channels probabilities $p_1$ and $p_2$, which were suggested by Eq.(3.8). The Hamiltonian of the pointer at atomic scale can be expressed as

$$\sum_k p_k^2/2m + \sum_{jk} V_{jk}(x_j - x_k), \qquad (4.8)$$

where the position and momentum observables of the constitutive atoms are denoted by ($x_k$, $p_k$). The potentials $V_{jk}$ ($j \neq k$) act between these particles and the summations range from 1 to $N$. When the distance $\xi$ is so small that proximity makes sense, the interactions $V_{jk}$ can act between an atom in a state $|x_1\rangle$ with its neighbors in a state $|x_2\rangle$ and not only in the state $|x_1\rangle$, the first case expressing the meaning of the coupling $H_{21}$ in Eq.(3.4).

Intuitively, the effect consists of transitions in which an atom or several ones transfer some probability from state $|x_1\rangle$ to state $|x_2\rangle$, without any change in the proper dynamics of

these two states. When seen in that way, the calculation looks difficult but, fortunately, it becomes much simpler when expressed in terms of exchanges among the probabilities of phonons. The method amounts to a canonical transformation from atomic observables to phonon observables. The convenience of this of this change was actually the main reason for considering the present model with a moving pointer, and phonons will be treated as colliding particles. This means that a phonon can be localized as a massless particle would be (see [31] for a discussion of this localization in the case of photons). This is possible when the phonon wave number is somewhat smaller than the size $d$ of a subsystem $\beta$ and one will consider only the effects arising fromthis category of phonons. The consideration of colliding particles allows also an estimate of the value of $d$ below which one cannot consider a wave function of phonons in a subsystem $\beta$ as approximately orthogonal to a wave function in a neighboring subsystem $\beta'$. This is based on multi-particle scattering and its accumulation of phase shifts, but would be too long to be developed here [32]. One will simply take the result as an assumption, namely that $d$ must be sufficiently larger than the average mean free path $\lambda$ of phonons.

An elementary process of probability fluctuations in a subsystem $\beta$ is then considered as a collision of two phonons $q$ and $q'$ in $\beta$ belonging to $|x_1\rangle$, yielding two phonons $q''$ and $q'''$ in $|x_2\rangle$ (or similar Umklapp processes). as well as opposite transitions from $|x_2\rangle$ to $|x_1\rangle$. The phonons $q$ and $q'$ carry then their probability from the matrix density $\rho_{\beta 1}$ to $\rho_{\beta 2}$ in Eq.(4.5) One computes in Appendix B the exchange of probability occurring during a small time $\delta t$. It arises from all the phonon pairs belonging to a coherent wave functions of phonons in $\beta$, and it is a random quantity with standard deviation

$$\delta p = (n_\beta^{-1} p_1 p_2 \delta t / \tau)^{1/2} \exp(-N_\beta \xi^2 / 8\alpha\Delta^2), \qquad (4.9)$$

where $n_\beta$ and $N_\beta$ denote respectively the number of phonons and the number of atoms in $\beta$. They are of the same order of magnitude when the temperature $T$ is of the order of the Debye temperature $\Theta$ or higher, namely $n_\beta \approx 3N_\beta(T/\Theta)$. In Eq.(4.9), $\tau$ is the mean free time of a phonon ($\tau = \lambda/c_s$, where $c_s$ is the sound velocity).The exponential in Eq.(4.9) is analogous to the expression (4.7) and equality, rather than equality holds because spectator phonons are identical in $|x_1\rangle$ and $|x_2\rangle$ (see Appendix B). The replacement of $N$ by $N_\beta$ reflects the incoherence of wave functions in different subsystems.

The process is not sensitive to the evolution of the density matrices $\rho_\beta$ resulting from the couplings $H_{\beta\beta'}$ in Eq.(4.6) between subsystems, because this evolution has a much longer time scale than phonon-phonon collisions. As shown also in Appendix B, the outcome is a random variation in the probabilities $p_1$ and $p_2$ of the two measurement channels during a time $\delta t$, with a standard deviation

$$\Delta \delta p_1 = (p_1 p_2 \delta t / \tau_{red})^{1/2}, \qquad (4.10)$$

and

$$1/\tau_{red} = 2(N/N_\beta^2)(\Theta/T)\exp[-N_\beta \xi^2 / 4\alpha\Delta^2](1/\tau). \qquad (4.11)$$

To get rough orders of magnitude, one may consider a cubic pointer with side $L$ and cubic subsystems with side $d$. Assuming the crystal lattice to have cubic symmetry and denoting the side of a lattice cell by $a$, one has $N/N_\beta^2 = (La/d^2)^3$. Taking for illustration $d = 10\,\lambda$, $\lambda = 3.\,10^{-7}$ cm (the mean free path of phonons in NaCl at room temperature [22]), $c_s = 3.\,10^5$ cm s$^{-1}$, $a = 3.\,10^{-8}$ cm, $L = 1$ cm, $\alpha = 1$, $T = \Theta$, $\Delta = 10^{-9}$ cm, one gets $\tau_{red} \approx 10^{-22} \exp(\xi^2/\xi_0^2)$ second with $\xi_0 \approx 10^{-11}$ cm. To evaluate for comparison the time during which proximity holds, one may notice that $\xi$ increases under the influence of a signal in $R$, which grows generally as an exponential $\exp(t/\tau')$. The time scale $\tau'$ of this growth is typically of the order $10^{-10}$ s and it controls the increase of $\xi$, irrespectively of the detailed mechanism for the action of the signal on the pointer. There is therefore plenty of time for reduction during the proximity period if $\tau_{red}$ is the time scale of reduction, as will be shown next.

*From fluctuations to reduction*

The next question asks whether these quantum fluctuations can yield reduction. Pearle gave the answer long ago [25], although he assumed that the fluctuations arise from nonlinear violations of the Schrödinger equation. Fortunately, his results do not depend on this assumption and can be used directly in the present framework. Pearle's conclusions can be summarized as a theorem, which deals with an arbitrary number of measurement channels, denoted by an index $j$, and it relies on the following assumptions:

*Assumption 1:* The probabilities $p_j$ of the various channels evolve randomly. One will call the resulting motion of the $p_j$'s a Brownian motion for convenience.

*Assumption 2:* The correlation functions $<\delta p_j \delta p_k>$ of the fluctuations occurring during a short time $\delta t$ are proportional to $\delta t$, as in a usual Brownian process. They depend only on time and on the $p_j$'s themselves. In the present case, this assumption is satisfied and one has actually

$$A_{jk} \equiv \langle \delta p_j \delta p_k \rangle / \delta t = -p_j p_k / \tau_{red} \qquad \text{for } j \neq k. \tag{4.12}$$

The relation $\sum_j \delta p_j = 0$ implies

$$A_{jj} \equiv \langle \delta p_j^2 \rangle / \delta t = p_j(1 - p_j)/\tau_{red}. \tag{4.13}$$

*Assumption 3:* If some probability happens to vanish during the Brownian motion, it remains zero afterwards. This is also valid in the present case, because no lost channel $k$ can be recovered ($p_k$ would have to start from 0 and the corresponding Brownian probability is proportional to $p_k$ and vanishes). Moreover, if a signal in the reactive part $R$ has disappeared through a loss of its amplitude, nothing can regenerate it.

According to Pearle's theorem, some channel probability $p_j$ must inevitably become equal to 1 after some time, randomly, and all the other probabilities vanish. Eq.(4.4) becomes then

$$\rho = \rho(R_j) \otimes \rho(P_j) \otimes |j\rangle\langle j|. \tag{4.14}$$

The first factor in the right-hand side shows that the signal corresponding to channel *j* is present in some reactive region, the second factor shows that the pointer position indicates the corresponding result and the last factor means that *m* is then in the pure state $|j\rangle$. Everything fits therefore wave function reduction, as it is usually defined when the measured system *m* is not destroyed and its state is not modified during the measurement.

An essential result of Pearle's theorem lies moreover in the predicted probability for this outcome of the Brownian process: It is exactly equal to the initial value of the quantity $p_j(0) = |c_j|^2$, i.e. in perfect agreement with Born's fundamental probability rule..

The proof of the theorem relies on a Fokker-Plank equation for a Brownian probability distribution $Q(p_1, p_2, p_3,...)$ describing the random variations of the the coefficients $p_k$ in Eq.(4.4). One has

$$\partial Q/\partial t = \sum_{jk} (\partial^2/\partial p_j \partial p_k) A_{jk} Q. \tag{4.15}$$

Many analyses of this equation by Pearle have shown that reduction is a random process ending after a random time *t* with a Poisson distribution $\exp(-t/\tau_{red})$ for large values of *t*. It can also be shown that, when there are many channels, the time of reduction does not depend appreciably on the number of channels and is always of the order of $\tau_{red}$ (there are simple geometric explanations for this result).

*Note 1: Probability distributions of probabilities.* One might question the meaning of a probability distribution *Q* for probabilities $p_k$. The answer is easy in the present case where $p_1$, for instance, is the quantity

$$Tr\{\rho.(I_S \otimes I_P \otimes I_R \otimes |1\rangle\langle 1|)\}. \tag{4.16}$$

The operators denoted by *I* are the identity operators for the various components of the apparatus and the projection operator $|1\rangle\langle 1|$ belongs to the measured system *m*, whereas $\rho$ is given by Eq.(4.4). The fluctuations in $p_1$ result from the quantum events in the pointer, which are independent of *m* and do not change the dynamics of the apparatus and of the channels. Because of this independence, Eq.(4.14) is no more surprising than taking into account random winds as a cause of variation in the probability that a bowman will hit his target. A nice point of quantum mechanics is on the other hand that it treats the wind, the bow and the arrow as a unique system, with identical basic rules and only takes account of organization.

*Note 2: Contagion of reduction.* When looking at the quantitative estimates in this section, one could question some values that were used for the parameters and ask for a more thorough review of various cases. This was not made because of a general property, which is already familiar in decoherence theory [12, 17], namely that many parts of an apparatus can work as pointers and this is also true of external phenomena, including for instance the reading of a datum by an experimenter. This contagion of decoherence is also valid for reduction and one may expect that complete reduction will occur somewhere, necessarily, whatever the different local values of $\tau_{red}^{-1}$, at least if they are not exceedingly large.

*Note 3: Perspectives of the present model.* The model is very restrictive since the example of a moving solid pointer is not representative of usual measuring devices. Another model where the pointer consists of an electric current has been also considered, the probability fluctuations being again due to phonons and their coupling with current arising from electron-phonon collisions. No final conclusion was yet ascertained however. Similarly, although Laughlin attributed to self-organization the origin of reduction and since droplets in a Wilson chamber or bubbles in a bubble chamber are self-organized, they were also considered from the present standpoint, but again with no definite result. An interesting possibility seems however to emerge, which would be a wide validity of incoherence in local wave functions and, correspondingly, an easy exchange of probability between self-organized superposed states. But this is at best a conjecture.

**5. Measurement of an EPR pair**

Quantum mechanics is non-separable [3], and this remarkable property has been confirmed beautifully by measurements of an EPR by two space-like separated apparatuses [26-28]. One might therefore wonder whether a theory relying on local reduction, like the present one, could agree with this type of results.

To examine this question, one considers two spin-1/2 particles 1 and 2 where the eigenvectors of the *z*-components of spin are entangled in a normalized state

$$|\psi\rangle = a|1z,+\rangle|2z,-\rangle + b|1z,-\rangle|2z,+\rangle.; \qquad (5.1)$$

where $|1z,+\rangle$, for instance, denotes the normalized state of particle 1 with its *z*-component of spin equal to +1/2, and the other notations are similar. Two space-like separated apparatuses $A_1$ and $A_2$ measure simultaneously the components of spin 1 and spin 2 along another space direction *z'* making an angle $\theta$ with the direction *z*. When expressed in the *z'*-basis, the state (5.1) becomes

$$|\psi\rangle = -(a+b)cs|1z',+\rangle|2z',+\rangle + (ac^2 - bs^2)|1z',+\rangle|2z',-\rangle$$
$$+ (bc^2 - as^2)|1z',-\rangle|2z',+\rangle + (a+b)cs|1z',-\rangle|2z',-\rangle, \qquad (5.2)$$

with $c = \cos(\vartheta/2)$, $s = \cos(\vartheta/2)$. For discussion, it will be more convenient to write down the four vectors in this expression as $|\alpha,\beta\rangle$ where the labels $\alpha$ and $\beta$ can take the two values ± and, for instance, $|++\rangle$ stands for $|1z',+\rangle|2z',+\rangle$. One has then

$$|\psi\rangle = \sum_{\alpha\beta} c_{\alpha\beta}|\alpha\beta\rangle. \qquad (5.3)$$

Experiments show that the observation of result $\alpha$ by $A_1$ and result $\beta$ by $A_2$ has the probability $p_\alpha = |c_{\alpha\beta}|^2$ or, in other words, the correlation coefficients are insensitive to the localization of measurements.

When applied to the present case and taking into account the independence of the two apparatuses $A_1$ and $A_2$, Eq.(4.1) becomes, at the beginning of measurements :

$$\rho_{R1+R2+m} = \sum_{\alpha\alpha'\beta\beta'} c_{\alpha\beta} c^*_{\alpha'\beta'} \rho(R_1)_{\alpha\alpha'} \otimes \rho(R_2)_{\beta\beta'} |1z',\alpha\rangle|2z',\beta\rangle\langle 1z',\alpha'|\langle 2z',\beta'|, \quad (5.4)$$

or after decoherence in the reactive regions,

$$\rho_{R1+R2+m} \approx \sum_{\alpha\beta} c_{\alpha\beta} c^*_{\alpha\beta} \rho^{(red)}(R_1)_{\alpha\alpha} \otimes \rho^{(red)}(R_2)_{\beta\beta} |\alpha\beta\rangle\langle\alpha\beta|. \quad (5.5)$$

Here, $\rho^{(red)}(R_1)_{\alpha\alpha}$ for instance, is the density matrix with trace 1 representing the signal in the reactive region $R_1$ of $A_1$ when the pure state $|\alpha\rangle$ is measured by $A_1$.

Let one then consider Pearle's Brownian reduction mechanism. Eq.(4.4) becomes

$$\rho_{A1+A2+m} \approx \sum_{\alpha\beta} p_{\alpha\beta} \{\rho(P_1,\alpha) \otimes \rho^{(red)}(R_1)_{\alpha\alpha}\} \quad A_{1,\alpha\beta,\alpha'\beta'} = \langle \delta p_{\alpha\beta 1}(t) \delta p_{\alpha'\beta'1}(t)\rangle$$
$$\otimes \{\rho(P_2,\beta) \otimes \rho^{(red)}(R_2)_{\beta\beta}\} \otimes (|\alpha\beta\rangle\langle\alpha\beta|). \quad (5.6)$$

Here $\rho(P_1,\alpha)$, with trace 1, represents for instance a state of the pointer $P_1$ in $A_1$, as it is moved by the signal generated by $\alpha$ in $R_1$.

The coefficients $p_{\alpha\beta}$ vary randomly with time and their fluctuations can be described exactly as in Section 4, except for taking into account that independent fluctuations occur in $A_1$ and $A_2$. This means that there are now different fluctuations $\delta p_{\alpha\beta 1}(t)$ of the coefficient $p_{\alpha\beta}$ arising from phonon effects in $A_1$, and independent fluctuations $\delta p_{\alpha\beta 2}(t)$ arising from effects in $A_2$. Because of independence, one has

$$A_{1,\alpha\beta,\alpha'\beta'} = \langle \delta p_{\alpha\beta 1}(t) \delta p_{\alpha'\beta'1}(t)\rangle, \quad (5.7a)$$

$$A_{2,\alpha\beta,\alpha'\beta'} = \langle \delta p_{\alpha\beta 2}(t) \delta p_{\alpha'\beta'2}(t)\rangle, \quad (5.7b)$$

$$\langle \delta p_{\alpha\beta 1}(t) \delta p_{\alpha'\beta'2}(t)\rangle = 0. \quad (5.7c)$$

One thus gets

$$A_{\alpha\beta,\alpha'\beta'} \equiv \langle \delta p_{\alpha\beta}(t) \delta p_{\alpha'\beta'}(t)\rangle = A_{1,\alpha\beta,\alpha'\beta'} + A_{2,\alpha\beta,\alpha'\beta'}, \quad (5.8)$$

and the random fluctuations of the quantities $p_{\alpha\beta}$ have exactly the same behavior as in Pearle's mechanism and their outcome is the same, namely that one of them becomes finally equal to 1, the other ones vanishing, and the Brownian probability for observing the final result $\alpha\beta$ is

$$p_{\alpha\beta}(t=0) \equiv |c_{\alpha\beta}|^2, \quad (5.9)$$

as one observed. Using the Fokker-Planck equation (4.15), an analogous analysis can be made when the two measurements are performed successively and also when the second one begins before a complete reduction of the first one, with identical results for the correlations.

## 6. Conclusion

A mechanism of reduction resulting from purely quantum effects has been found efficient in a special model of measurement. Counter-arguments based on no-go theorems were also questioned and the hypothesis of consistency between the quantum principles and reduction can be accordingly envisioned as possible. Because of special features of the model and of some approximation that had to be made, this is proposed as a conjecture requiring more investigation. If this conjecture were shown to be true however, the uniqueness of reality would be an emergent property with a specific domain of validity, like classical physics. It would also provide a remarkable example of self-consistency within the quantum principles and of their ability to generate the concepts for their own interpretation

## Acknowledgements

Many friends and colleagues contributed to this work by useful criticisms, remarks, questions and suggestions. Encouragements by Robert Dautray, Jacques Friedel and Phillip Pearle were precious. I thank particularly Bernard d'Espagnat for his thorough critique over years and his invaluable help when stating the non-separability problem and contributing personally to its answer.

## Appendix A

## About wave functions

The most important requirement for a model of reduction is probably to understand how it can circumvent the negative results or no-go theorems excluding consistency between reduction and Schrödinger's dynamics [3-6]. Although not pretending that these results can be disproved, one will point out here some apparently unnoticed questions, which are suggested by the present approach.

These works relied strongly on the use of wave functions, sometimes extended to a large part of the universe or to the whole universe itself [6], and also sometimes extended to statistical ensembles of such functions [5], amounting to a density matrix. A well-known difficulty is however the physical definition of these fundamental concepts [2]. Given for instance a density matrix, one might define the relevant wave functions as its eigenfunctions, but how can one define the density matrix itself? To say that it is a partial trace of a density matrix for a larger system is a circular argument, leading immediately to introduce the universe. This is why one is constrained more or less to assume that the measuring device is isolated. This assumption, however, does not tell one directly what a relevant wave function should be, except if it is an eigenfunctions of a given density matrix. But then, fortunately, one can rely on Gleason's theorem asserting under simple and sensible assumptions that a density matrix exists [Gleason], although one cannot say exactly what it is. Accordingly, one will assume in this appendix that the systems under consideration is isolated and has a well-defined density matrix $\rho$ so that, when a wave function $\psi$ is considered, it is an eigenfunction of $\rho$.

The analysis in Section 4 suggested that an approximate factorization of $\rho$, accounting for organization and for the notion of Gibbs ensembles, could shed some light on reduction. In the framework of a fundamental approach however, one must be careful about the use of decoherence [2, 29], and in place of Eq.(4.4), organization would lead only to the approximation:

$$\rho \approx \sum_{i,j} \rho_{ij}(R) \otimes \rho_{ij}(P) \otimes |i\rangle\langle j| ., \quad (A.1)$$

the summation being carried over the values 1 and 2 for the indices $i$ and $j$. This factorization, if it makes sense, would imply that the eigenfunctions of $\rho$ are themselves approximately factorized so that the main aim of this appendix will be to consider the meaning and consequences of this kind of approximate factorization.

To begin with, one considers a normalized wave function $\psi(x_1, x_2)$ depending on two variables. One can always introduce a basis of orthogonal function $\varphi_n(x)$ for the two variables and write

$$\psi(x_1, x_2) = \sum_{nn'} c_{nn'} \varphi_n(x_1) \varphi_{n'}(x_2), \quad (A.2)$$

which is a sum of factorized quantities. But in general, no coefficient $c_{nn'}$ dominates this sum and a small number of terms does not represent correctly the function $\psi(x_1, x_2)$. The problem under consideration is different and asks whether a better choice of basis, depending on $\psi$, can allow a better physical significance for a finite number of components..

One considers therefore the question: What are the two functions $\varphi_1(x_1)$ and $\varphi_2(x_2)$ such that the square norm $J = \|\psi - \varphi_1 \varphi_2\|^2$ has a minimum? One may consider first for convenience the case of real functions and, to remove ambiguities in the product $\varphi_1(x_1) \varphi_2(x_2)$, one takes a normalized factor $\varphi_1(x_1)$:

$$\|\varphi_1\|^2 = \int \varphi_1^2(x_1) dx_1 = 1. \quad (A.3)$$

One also writes down

$$\|\varphi_2\|^2 = \mu, \quad (A.4)$$

where $\mu$ is a parameter to be determined. Writing down

$$J = \int\int \{\psi(x_1, x_2) - \varphi_1(x_1)\varphi_2(x_2)\}^2 dx_1 dx_2 , \quad (A.5)$$

Performing variations $\delta\varphi$ of the $\varphi$ functions, one gets, after introducing a Lagrange coefficient $\lambda$ to take condition (A.3) into account:

$$\int \psi(x_1, x_2) \varphi_1(x_1) dx_1 = \varphi_2(x_2), \quad (A.6)$$

$$\int \psi(x_1,x_2)\varphi_2(x_2)dx_2 = (\lambda-\mu)\varphi_1(x_1). \quad (A.7)$$

Using Eq.(A.6), Eq.(A.7) becomes an eigenvalue integral equation

$$K\varphi_1 = (\lambda-\mu)\varphi_1, \quad (A.8)$$

for the integral operator $K$ with kernel

$$K(x_1,x'_1) = \int \psi(x_1,x_2)\psi(x'_1,x_2)dx_2 \quad (A.9)$$

$K$ is a positive operator with trace 1 and, if one assumes furthermore that $\psi$ is continuous and the domains of the two variables $x_1$ and $x_2$ are finite intervals, $K$ is also a compact operator and must have discrete positive eigenvalues. Let then $\nu$ denote the largest eigenvalue, assume it non-degenerate and choose the corresponding normalized eigenfunction for $\varphi_1(x_1)$. One gets then the relations: $\mu = \nu$, $\lambda = 2\nu$, and

$$\varphi_2(x_2) = \int \psi(x_1,x_2)\varphi_1(x_1)dx_1. \quad (A.10)$$

One checks easily that the same result is obtained when the constraint is a normalization of $\varphi_2$ rather than $\varphi_1$.

The next step consists in considering the function $\psi(x_1,x_2) - \varphi_1(x_1)\varphi_2(x_2)$ in place of $\psi(x_1,x_2)$. Straightforward calculations show then that the best approximation is given by

$$\varphi_1\varphi_2 + \varphi_1^{(1)}\varphi_2^{(1)}, \quad (A.11)$$

where $\varphi_1^{(1)}$ is the normalized eigenfunction of $K$ with the next largest eigenvalue and $\varphi_2^{(1)}$ is related to it by Eq.(A.6). This procedure can be pursued to obtain more and more factorized components and the case of degenerate eigenvalues implies no special difficulty. The case of complex functions is practically identical and yields similar results.

In the case of three variables $(x_1,x_2,x_3)$, a remarkable property comes out with the occurrence of nonlinear features. When one tries to apply directly the same method to a product of three functions, the variational approach involves quadratic integrals such as

$$\int f(x_1,x_2,x_3)\varphi_1(x_1)\varphi_2(x_2)dx_1dx_2 \quad (A.12)$$

and one gets a set of nonlinear integral equations in place of Eqs.(A..6,7). One may try then to proceed step by step, using the previous method with two functions $\varphi_1(x_1,x_3)$ and $\varphi_2(x_2,x_3)$ depending on $x_3$. One can then approximate $\varphi_1(x_1,x_3)$ by a sum with a first term $\varphi_{11}(x_1)\varphi_{13}(x_3)$ and similarly for $\varphi_2(x_2,x_3)$. Replacing the notations $\varphi_{11}$ and $\varphi_{22}$ by $\chi_1$ and $\chi_2$, and putting $\chi_3 = \varphi_{13}\varphi_{23}$, one gets a leading factor $\chi_1\chi_2\chi_3$ in the expression of $\psi(x_1,x_2,x_3)$. The result depends however on the order in which the set $(x_1,x_2,x_3)$ is split into two selected variables ($x_1$ and $x_2$ in this example) and an outside variable ($x_3$). There is no mathematical reason for the three different selections of variables to yield the same result and that means necessarily that another product exists and provides generally a better approximation, signifying again nonlinearity.

It may be noticed that these method do not assume that $x_1$, for instance, is a unique variable. The same results are valid when $x_1$ stands for all the coordinates of the particles in a piece of an organized system and $x_2$ similar coordinates in another piece. The main point is that factorization is very sensitive to the basic wave function $\psi(x_1,x_2)$: It may be expected much more efficient in the case of an organized system than for a generic (arbitrary) mathematical wave function, in which case there is presumably no sensible factorization.

These simple remarks draw attention to a missing opportunity in all the works in which a wave function $\psi$ is used to represent a measuring device [2-6]. Except for introducing an observable representing the pointer position, nothing more is said or used about $\psi$. But this wave function is far from being arbitrary. It has always organization and a past history, responsible for instance for its thermal properties. Organization is certainly the most significant feature, since it constrains strongly the wave functions, as shown in Section 2, and one may remember that most quantum calculations using boundary conditions for a subsystem make use of these constraints. The discussion in Section 4, when applied to organized pieces of an apparatus, has suggested that interactions can take place between two different components of $\psi$, represented by different terms in a sum of factors and dealing with the same microscopic observables.

There are always many pieces in an actual measuring device, or many factors in a component of $\psi$. If a reduction mechanism such as the one in Section 4 is valid, it can take place between two factors for the same piece, for instance $\varphi(x_1)$, entangled with $|1\rangle$ (which is one factor) and $\varphi'(x_1)$, entangled with $|2\rangle$, and this possibility is of course the main effect that should be reconsidered in the fundamental approach. It does not matter then that the sum of factorized components does not represent $\psi$ exactly and it is enough that it represents a significant part of $\psi$.

Finally, I suggest that no-go assertions should be reconsidered and might perhaps be revised. An obvious difficulty for making this analysis is however the nonlinearity of factorization and its strong dependence on $\psi$, together with the fact that it changes when $\psi$ evolves. In that sense, the fundamental approach using wave functions looks more difficult than a phenomenological approach, which is better suited to the actual physics of a system. Moreover, one may expect that factorization makes sense since physics has always made a wide use of intuitive, or sensible, or partly rigorous properties of factorization in statistical physics, solid-state physics, or simply when one relies on boundary conditions, and many resulting predictions agreed with experiment.

**Appendix B**

**Fluctuations of channel probabilities**

When considering scalar products for the states of a pointer or of one of its subsystems, it will be convenient to use a model where $N$ atoms are bound to their lattice sites by harmonic potentials with frequency $\omega$. One denotes by $\Delta$ the standard deviation of an atom position in the ground state whereas in an excited state with energy $(n+1/2)\hbar\omega$, this uncertainty is given by $\Delta_n^2 = (2n+1)\Delta^2$. At zero temperature, the atoms are in their ground states $|x_1\rangle$ and $|x_2\rangle$ with Gaussian wave functions and one has

$$\langle x_1|x_2\rangle = \exp(-N\xi^2/8\Delta^2) \tag{B.1}$$

with $\xi = x_1 - x_2$. At a finite temperature $T$, the average displacement of an atom is given by $\langle \Delta_n^2\rangle = \alpha\Delta^2$, $\alpha = (\gamma+1)/(\gamma-1)$, $\gamma = \exp(\hbar\omega/T)$ and one will estimate the average scalar product of two identical states, relatively displaced by a translation $\xi$, as given approximately by

$$\langle x_1|x_2\rangle = \exp(-N\xi^2/8\alpha\Delta^2) \tag{B.2}$$

When the system under consideration is a subsystem $\beta$ of a larger system ($N \to N_\beta$), its position fluctuates with respect of the center of mass of the large system with a standard deviation $\Delta'$ ($\Delta'^2 = (\alpha-1)\Delta^2/N_\beta$) and one sees that these global displacements do not affect significantly the value of $\xi$ in the exponential of Eq.(B.2). In other words, one can neglect the fluctuations of the position of $\beta$.

One will now consider phonons, as in Section 4, and introduce some notations. The vector wave number of a phonon is denoted by $q$ and the corresponding occupation number by $n(q)$. The polarization of phonons will not be written down explicitly, because it is inessential in the present considerations. A phonon state is denoted by $|\{n(q)\}\rangle$ and the various phonon states are incoherent. A tedious straightforward calculation yields, at a finite temperature::

$$\langle\{n_1(q_k)\},x_1|\{n_2(q_k)\},x_2\rangle \approx \prod_k \delta[n_1(q_k),n_2(q_k)].\exp(-N\xi^2/8\alpha\Delta^2). \tag{B.3}$$

The considerations from statistical physics, which lead to Eq.(4.5) imply that the states of two different subsystems $\beta$ and $\beta'$ are incoherent. One may wonder however what is the distance at which correlations become significant and cannot be neglected. One will assume that it is of the order of the mean free path of phonons $\lambda$ and, although this assumption is suggested by multi-particle scattering [32], it should be considered as one of the main approximations in the present calculations.

One considers now phonon collisions in a cell $\beta$ with size $d$ larger than $\lambda$ but of the same order of magnitude. The pointer is first considered in its initial state before measurement or during measurement when a pure state $|j\rangle\langle j|$ is measured. The common point in these various cases is that the pointer position is well defined and unique at a given time. A phonon can be considered as localized in $\beta$ as long as $q\lambda > 1$ and only this category of phonons will be considered [31]. The Hamiltonian $H_\beta$ in Eq.(4.6) involves a sum of phonon energies and their interaction yielding phonon collisions. A scattering between two phonons with wave numbers $q$ and $q'$ and yielding two phonons $q''$ and $q'''$ is due for instance to a coupling term

$$V = \sum C(q,q';q'',q''')a^\dagger(q'')a^\dagger(q''')a(q)a(q'), \qquad (B.4)$$

where $a^\dagger$ and $a$ denote creation and annihilation operators. There are also Umklapp processes corresponding for instance to a decay $q \to q' + q'' + q'''$ and three-phonons interactions yielding Umklapp processes and no scattering, but their discussion is similar and will not be developed. Because $d \ll L$, most state vectors $|\{n_\beta(q)\}\rangle$ for the phonons located in $\beta$ have occupation numbers $n_\beta(q)$ equal to 0 or 1, most of them taking the value 0. When considering a transition between different states of $\beta$, a scattering $q + q' \to q'' + q'''$ is a transition

$$|q,q',s\rangle \to |q'',q''',s\rangle, \qquad (B.5)$$

where $s$ stands for the spectator phonons in $\beta$, which do not participate in the reaction. They must be identical in the initial and the final sate. One usually omits these spectators in solid-state physics where they play no role, but they must be written down here, because their role will become significant when superposed states of the pointer will be considered. The scattering amplitude is governed by a matrix element $\langle q'',q''',s|V|q,q',s\rangle$ with conservation of energy and momentum, yielding after a summation of its squared modulus over $q''$, $q'''$ and over the states $s$ a probability $w$ per unit time for a collision between $q$ and $q'$ in $\beta$, with $w = (n_\beta \tau)^{-1}$ on average.

One comes now to the crucial case when a state $|\psi\rangle\langle\psi|$ of $m$ is measured and the two possible states of the pointer are distant of $\xi$. There are then different kinds of states in $\beta$, which differ by the state $|j\rangle$ with which they are entangled and they can be written respectively as $|\{n_{\beta 1}\},x_1\rangle|1\rangle$ and $|\{n_{\beta 2}\},x_2\rangle|2\rangle$. Corresponding states of the reactive region $R$ are supposed included in the notation $|1\rangle$ or $|2\rangle$ for shorter writing. Eq (B.3) becomes then

$$\langle\{n_{\beta 1}(q_k)\},x_1|\{n_{\beta 2}(q_k)\},x_2\rangle \approx \prod_k \delta[n_{\beta 1}(q_k),n_{\beta 2}(q_k)]e', \qquad (B.6)$$

with

$$e' = \exp(-N_\beta \xi^2/8\alpha\Delta^2). \qquad (B.7)$$

This replacement of the total number of atoms $N$ by the much smaller number $N_\beta$ in $\beta$ ($N_\beta \approx (L/d)^3 N$) is striking, because it will also occur in the final result: It reflects of course the limited range of correlations in wave functions, but in a rather spectacular way.

One can consider again a phonon collision $q + q' \to q'' + q'''$. but taking now place between the vectors $|q,q',s\rangle|j\rangle$ and $|q'',q''',s\rangle|k\rangle$ where $j$ and $k$ equal to either 1 or 2 and the spectator $s$ are identical. The corresponding collision matrix element is now

$$\langle q'',q''',s,x_k|Ca^\dagger(q'')a^\dagger(q''')a(q)a(q')|q,q',s,x_j\rangle. \qquad (B.8)$$

It coincides with the previous matrix element yielding the transition (B.5) when $q$, $q'$, $q''$ and $q'''$ are associated with the same $j$ ($j = k$). When $j \neq k$ however, it is multiplied by the factor $e'$. The probability for a transition of two phonons ($q, q'$) both entangled with $|1\rangle$, yielding ($q'', q'''$), both entangled with $|2\rangle$ is obtained after summing over all the spectators. It involves the same collision rate as in a unique pointer state when either the $m$-state $|1\rangle$ or $|2\rangle$ was measured), but it is now multiplied by the very significant factor $p_1 p_2 . e'^2$. $p_1$ and $p_2$ are the probabilities of the $m$-states $|1\rangle$ and $|2\rangle$, and their occurrence is due to the quantum amplitude factors $p_1^{1/2}$ and $p_2^{1/2}$ of the spectator in $\beta 1$ and $\beta 2$, and the factor $e'$ is their scalar product. There are $n_\beta$ phonons entangled with $|1\rangle$, each one of them can only scatter with another phonon entangled with $|1\rangle$. Since there are $n_\beta^2/2$ pairs of phonons and every collision with transfer of phonons brings two phonons from $|1\rangle$ to $|2\rangle$ (or the other way round), the total number of phonons going from $|1\rangle$ to $|2\rangle$ during a short time $\delta t$ is $(n_\beta \delta t/\tau) p_1 p_2 e'^2$. Each phonon that is entangled with $|1\rangle$ brings with it on average its probability $n_\beta^{-1}$. Since there are on average as many transitions in both directions with transfers of probabilities with different signs, the resulting fluctuations in $p_1$ and $p_2$ during a time $\delta t$ are random with standard deviation

$$(n_\beta^{-1} p_1 p_2 e'^2 \delta t/\tau)^{1/2} \tag{B.9}$$

The next step consists in going from the fluctuations of the probabilities $p_1$ and $p_2$ in a subsystem $\beta$ to fluctuations in the whole pointer and it relies strongly on the expression (4.5) for the density matrix of the system $A + m$, where all the matrices $\rho_{\beta 1}$ and $\rho_{\beta 2}$ have trace 1. After a time $\delta t$, phonon collisions transform $\rho_{\beta 1}$ into another matrix $\rho'_{\beta 1}$ and $\rho_{\beta 2}$ in $\rho'_{\beta 2}$. This transition is complex, but one is only interested in the variations of their traces resulting from local exchanges in probability. If a probability $\delta p_\beta$ with either sign is carried during the time $\delta t$ from $\beta 1$ to $\beta 2$, the trace of $\rho'_{\beta 1}$ becomes $1 + \delta p_\beta$ and the trace of $\rho'_{\beta 2}$ becomes $1 - \delta p_\beta$. One has then

$$Tr(\prod_\beta \rho'_{\beta 1}) = \prod_\beta (1 + \delta p_\beta) \approx 1 + \sum_\beta \delta p_\beta, \tag{B.10}$$

where the last equality results from the smallness of the quantities $\delta p_\beta$. At time $t + \delta t$, Eq.(4.5) becomes

$$\rho' = p_1 \left(\prod_\beta \rho'_{\beta 1}\right) \rho(R_1) |1\rangle\langle 1| + p_2 \left(\prod_\beta \rho'_{\beta 2}\right) \rho(R_2) |2\rangle\langle 2|. \tag{B.11}$$

It can be rewritten in the previous form (4.5) involving products of matrices with unit trace after a renormalization $\rho'_{\beta 1} = (1 + \delta p_\beta) \rho''_{\beta 1}$, $\rho'_{\beta 2} = (1 - \delta p_\beta) \rho''_{\beta 2}$, so that the traces of the matrices $\rho''_{\beta j}$ are equal to 1. One thus gets

$$\rho' = p'_1 \left(\prod_\beta \rho''_{\beta 1}\right) \rho(R_1) |1\rangle\langle 1| + p'_2 \left(\prod_\beta \rho''_{\beta 2}\right) \rho(R_2) |2\rangle\langle 2|, \tag{B.12}$$

which has still the form (4.5), but with different values of the channel probabilities $p'_1$ and $p'_2$ after a time $\delta t$. One has

$$p'_1 = p_1[1+\sum_\beta \delta p_\beta], \; p'_2 = p_2[1-\sum_\beta \delta p_\beta],$$

and hence

$$p'_1 = p_1 + (1/2)\sum_\beta \delta p_\beta, \; p'_2 = p_2 - 1/2)\sum_\beta \delta p_\beta. \qquad (B.13)$$

Since the number of subsystems $\beta$ is equal to $N/N_\beta$, one obtains Eqs.(4.10-11) for the standard deviation of probability fluctuations during a time $\delta t$.

**Notes and References**